\documentstyle[pra,aps,multicol]{revtex}
\begin{document}

\draft
\title{The role of imaginary vector potential 
in composite fermion pairing theory}
\author{Takao Morinari}
\address{Department of Applied Physics, University of Tokyo,
Bunkyo-ku, Tokyo 113-8656, Japan}

\date{\today}
\maketitle
\begin{abstract}
We show that the imaginary vector potential causes a pair-breaking effect
in the composite fermion theory, and discuss its irrelevance in the pairing
state. The Hamiltonian for pairing states of composite fermions is proposed.
The advantage of non-unitary transformations and the meaning of the
non-Hermitian term are discussed.
\end{abstract}
\pacs{73.40.Hm, 71.10.Pm}

\begin{multicols}{2}
\narrowtext
Recently, the $\nu=5/2$ enigma is reconsidered in Refs.
\cite{morf,rezayi_haldane,park}.
These numerical works support that the $\nu=5/2$ state is the spin-polarized
p-wave pairing state of composite fermions\cite{jain}.
On the other hand, the possibility of the p-wave pairing state was
discussed at $\nu=1/2$ \cite{gww,morinari,park}.

However, the condition for the occurrence of composite fermion pairings
are still controversial problem.
The pair-breaking effect in the composite fermion
pairing state was discussed by Bonesteel \cite{bonesteel}.
He discussed that there is a pair-breaking effect stemmed from 
a fluctuation of the Chern-Simons gauge field and
this pair-breaking effect is strong enough in a short range interaction case
to break composite fermion pairs.
On the other hand,
the effect of non-Hermitian term was unclear in the composite fermion 
pairing theory \cite{morinari} based on the Rajaraman-Sondhi transformation 
\cite{rajaraman_sondhi}.

The effect of non-Hermitian term was discussed in 
the localization-delocalization phenomena\cite{hatano_nelson}.
For two-dimensional non-interacting electron systems with 
random potentials,
it is believed that all states are localized\cite{localization}.
However, when we introduce an imaginary vector potential, 
and increase the strength
of it, there appear extended states\cite{hatano_nelson}.
The origin of the imaginary vector potential is clear for the 
vortex depinning phenomena with columnar defects,
that is, the transverse magnetic field\cite{hatano_nelson}.
However, the merits of considering the non-Hermitian quantum mechanics 
in general systems are not clear.

In this paper, we show that the non-Hermitian term causes a pair-breaking
effect to composite fermion pairing states.
By solving the gap equations, it is shown 
that the ground state is the gapless 
pairing state in the absence of the Coulomb energy. 
However, this gapless pairing state is unstable one because it is 
on the relative maximum of the ground state energy. 
When we take into account the Coulomb interaction, it is not the ground state
any more.
The advantage of non-unitary transformation and the meaning of the 
non-Hermitian term are discussed. Moreover we mention a guide to apply 
non-unitary transformations to the strongly correlated electron systems.

The Hamiltonian at $\nu=1/m$ ($m$ is an even integer)
for composite fermions based on the Rajaraman-Sondhi
non-unitary transformation \cite{rajaraman_sondhi} 
is given by \cite{morinari} 
\begin{equation}
H = H^0 + V^H + V^{NH} + V^C,
\label{rs_h}
\end{equation} 
where
$H^0 = \sum_{\bf k} \frac{\hbar^2 k^2}{2m_b} \pi_{\bf k} \phi_{\bf k}$,
$V^C$ is the Coulomb interaction, and
\begin{eqnarray}
V^H & = & \int d^2 {\bf r}~ \frac{e}{c}~
{\bf j}_{CF} ({\bf r}) \cdot \delta {\bf a},
\label{v_h} \\
V^{NH} & = & \int d^2 {\bf r}~ \frac{e}{c}~
{\bf j}_{CF} ({\bf r}) \cdot \left( i~ \hat{e}_z \times \delta {\bf a} \right),
\label{v_nh}
\end{eqnarray}
with $\pi_{\bf k}$ ($\phi_{\bf k}$) being the creation (annihilation) operator 
of composite fermions with momentum ${\bf k}$, 
${\bf j}_{CF}$ being the current operator of composite
fermions, and $\delta {\bf a}$ being given by
\begin{equation}
\delta {\bf a}_{\alpha}
= \frac{\phi_0}{2\pi}~ m \int d^2 {\bf r}^{\prime} 
  \delta \rho_{CF}({\bf r}^{\prime})
\nabla {\rm Im} \log (z-z^{\prime}).
\label{cs_g}
\end{equation}
Here $\phi_0 = \frac{ch}{e}$ is the flux quantum, $z=x+iy$ is the complex
coordinate, and $\delta \rho_{CF}({\bf r})$ is equal to
$\rho_{CF} ({\bf r}) - \overline{\rho}$ with $\overline{\rho}$ being
the average density of particles.
Substituting Eq. (\ref{cs_g}) into Eqs. (\ref{v_h}) and (\ref{v_nh}) directly,
we obtain the interaction between composite fermions in two-body forms.

Equation(\ref{v_h}) has the form of the minimal coupling 
between composite fermions and the Chern-Simons gauge field fluctuation.
Therefore, $V^H$ causes an interaction like the Lorentz force.
Considering the classical equation of motion for composite fermions,
we see that a composite fermion passing by another composite fermion
in the counterclockwise direction from the view point at positive $z$-axis
feels the attractive force toward it because
$\delta \rho >0$ around the composite particles.
On the other hand, if a composite fermion passes by another composite fermion
in the clockwise direction, the repulsive interaction is caused between them.
For that reason, a pairing state with positive angular momentum is 
expected to exist\cite{pairing}. From this argument, we expect that if there
is some pairing state, the symmetry of the pairing state is not s wave.
On the other hand, we find that the $V^{NH}$ has no effect for the equation
of motion.

The gap equations for pairing states of composite fermions are derived
by a pairing approximation. At zero temperature, they are given by
\begin{eqnarray}
\Delta_{\bf k} & = & - \frac{1}{2\Omega} \sum_{{\bf k}^{\prime} (\neq {\bf k})}
V_{{\bf k} {\bf k}^{\prime}} 
\frac{\Delta_{{\bf k}^{\prime}}}{E_{{\bf k}^{\prime}}},
\label{gap} \\
\overline{\Delta}_{\bf k} 
& = &  - \frac{1}{2\Omega} \sum_{{\bf k}^{\prime} (\neq {\bf k})}
V_{{\bf k}^{\prime} {\bf k}} \frac{\overline{\Delta}_{{\bf k}^{\prime}}}
{E_{{\bf k}^{\prime}}},
\label{gapbar}
\end{eqnarray}
where $V_{{\bf k} {\bf k}^{\prime}}$ is the interaction for pairs with 
zero total momentum.
The analysis of these gap equations with $V^H$ only for the pairing interaction
was done by Greiter, Wen, and Wilczek \cite{gww}.
They showed that the ground state is 
the p-wave pairing state of composite fermions.
In their analysis, the usual Chern-Simons singular gauge transformation was
performed. The quadratic term for the Chern-Simons gauge field was neglected
and the effect of it was not clear.
As we will show later, it has close relation to 
the non-Hermitian term $V^{NH}$.

Now we take into account the non-Hermitian term $V^{NH}$ in the gap equations.
Setting $\Delta_{\bf k} = \Delta_k {\rm e}^{-i \ell \theta_k}$
and 
$\overline{\Delta}_{\bf k} = \overline{\Delta}_k {\rm e}^{i \ell \theta_k}$
in Eqs. (\ref{gap}) and (\ref{gapbar})
for $\ell$-pairing state ($\tan \theta_k= \frac{k_x}{k_y}$), we obtain
\begin{eqnarray}
\Delta_k 
& = & \frac{m}{M} \int_0^k d k^{\prime} \frac{k^{\prime} 
\Delta_{k^{\prime}}}{E_{k^{\prime}}}
\left( \frac{k^{\prime}}{k} \right)^{\ell},\\
\overline{\Delta}_k 
& = & \frac{m}{M} \int_k^{\infty} d k^{\prime} \frac{k^{\prime} \Delta_{k^{\prime}}}{E_{k^{\prime}}}
\left( \frac{k}{k^{\prime}} \right)^{\ell},
\end{eqnarray}
where we have replaced the band mass with the effective mass $M$ 
of composite fermions.
These gap equations are solved exactly and the solution is given by
\begin{eqnarray}
\Delta_k 
& = & \left\{ \begin{array}{lc}
0 & ~\bbox{for}~k<k_F, \\
\Delta \epsilon_F \left[ \left( \frac{k_F}{k} \right)^2 -1\right]^m
\left( \frac{k}{k_F} \right)^{\ell} &
~\bbox{for}~k>k_F,
\end{array}
\right.\\
\overline{\Delta}_k
& = & \left\{
\begin{array}{lc}
\overline{\Delta} \epsilon_F \left(\frac{k}{k_F}\right)^{\ell} 
\left[ 1-\left(\frac{k}{k_F}\right)^2\right]^m &
~\bbox{for}~k<k_F, \\ 
0 & ~\bbox{for}~k>k_F, 
\end{array}
\right.
\end{eqnarray}
where $\Delta$ and $\overline{\Delta}$ are constants, $\epsilon_F$ and
$k_F$ are the Fermi energy and the Fermi wave vector for composite fermions,
respectively.
As a remarkable fact, this state is the gapless pairing state 
because $\overline{\Delta}_k \Delta_k \equiv 0$ for any $k$ but
each of $\overline{\Delta}_k$ and $\Delta_k$ is not identical to zero.
Therefore, we understand that the effect of $V^{NH}$ is the 
pair-breaking effect. 
The conclusion that all of the pairing states are gapless
if we neglect the Coulomb interaction is the natural one.
In the absence of the Coulomb interaction, the original problem
is the free electron gas under the magnetic field.
In this situation, it is not an appropriate starting point to take an 
approximation to the Hamiltonian obtained by either the Chern-Simons 
singular gauge transformation or Rajaraman-Sondhi transformation
because the two-body correlation effect, which is taken into account 
by such transformations, is absent.
We do not expect the pairing correlation in such systems.

The pair-breaking effect caused by $V^{NH}$ has a close relation 
to the effect of an 
imaginary vector potential in the localization-delocalization phenomena.
From Eq. (\ref{v_nh}), we see that $V^{NH}$ corresponds to an imaginary 
vector potential $i \hat{e}_z \times \delta {\bf a}$.
The effect of the imaginary vector potential was discussed in the 
localization-delocalization phenomena\cite{hatano_nelson}.
In the absence of the imaginary vector potential, it is believed that
all eigenstates are localized in two-dimensional non-interacting 
systems\cite{localization}.
However, if we introduce an imaginary vector potential, and
increase the strength of it, there appear extended states\cite{hatano_nelson}.
In the composite fermion pairing theory, as we have seen above,
the ground state of the system
is the pairing state in the absence of the imaginary vector potential
and the Coulomb interaction.
When we take into account the imaginary vector potential, which 
is fixed contrary to the vortex depinning phenomena \cite{hatano_nelson}, 
the gap of the pairing state goes to zero.
Therefore, the imaginary vector potential in the composite
fermion pairing state has the similar effect as in 
the localization-delocalization phenomena.

As discussed above, the ground state is the gapless pairing state 
in the absence of the Coulomb interaction.
However, this gapless pairing state is not stable 
because, from discussion below, we see that it is on the relative maximum 
of the ground state energy.
The variation of the ground state energy with regard to 
the $\overline{\Delta}_k \Delta_k$ is given by
\begin{eqnarray}
\langle H \rangle_{\overline{\Delta}_{k} \Delta_{k} + 
\delta \left( \overline{\Delta}_{k} \Delta_{k} \right)}
& - &
\langle H \rangle_{\overline{\Delta}_{k} \Delta_{k}} \nonumber \\
&   & \hspace{-20mm} = - \frac{1}{8} \sum_{\bf k} \frac{\overline{\Delta}_{k} \Delta_{k}}
{\left(\overline{\Delta}_{k} \Delta_{k} + \xi_k^2 \right)^{3/2}}
~\delta \left( \overline{\Delta}_{k} \Delta_{k} \right).
\end{eqnarray}
The factor
$ \overline{\Delta}_{k} \Delta_{k}/
\left(\overline{\Delta}_{k} \Delta_{k} + \xi_k^2 \right)^{3/2}$
is not lower than zero.
Therefore, the function  $\overline{\Delta}_{k} \Delta_{k} \equiv 0$ is 
the relative maximum of $\langle H \rangle$.
It means that any perturbation will unstabilize the gapless pairing state.
Hence when we take into account the Coulomb interaction, this gapless pairing
state is not stable.
The effect of $V^{NH}$ is the pair-breaking effect but it is irrelevant 
for the pairing state.

The irrelevance of $V^{NH}$ is understood by considering the meaning of 
the Rajaraman-Sondhi non-unitary transformation.
It fully takes into account the most fundamental correlation in quantum 
Hall systems.
Because of the Coulomb interaction between electrons and the external magnetic
field, any pair of electrons has the correlation with non-zero relative angular
momentum as the short-range correlation.
The Rajaraman-Sondhi transformation fully takes account of it.
Then, what is the meaning of a non-Hermitian term resulting from
the Rajaraman-Sondhi transformation?
Before discussing it, we consider the simplest problem.
Suppose the harmonic oscillator in one dimension:
$H= - \frac12 \frac{d^2}{dx^2} + \frac12 x^2$.
As is well-known, the ground state is $\psi_0(x) \propto \exp(-\frac{x^2}{2})$.
Let us apply a non-unitary transformation:$U=\psi_0(x)$ via 
$\overline{U}HU$, where $\overline{U}=\exp(\frac{x^2}{2})$.
Of course the resulting Hamiltonian has the non-Hermitian term.
However, it only affects excited states. For the ground state, it is not
relevant.
Similarly, the effect of $V^{NH}$ is only relevant for the motion
which deviates from the above fundamental correlation.
However, such motions are high energy modes in the presence of the 
Coulomb interaction.
If it is relevant, the Hamiltonian obtained by the Rajaraman-Sondhi 
transformation as well as the Chern-Simons singular gauge transformation is not
effective one.
As far as the above correlation for short range correlation is 
the most fundamental one, $V^{NH}$ is irrelevant.
The fact that the above correlation effect is the most fundamental one
is demonstrated for the bilayer quantum Hall systems\cite{morinari_b}.

From the discussion above, we get a guide for the application of
non-unitary transformations.
The correlation effect in condensed matter physics is divided into two parts:
short range correlation effects and long range correlation effects.
The difficulty of dealing with strongly correlated electron systems 
theoretically relies on that we do not know how to take into account 
the former.
However, the short range correlation originates from 
the two-body correlation, 
and it is found by considering the two-body problem.
Therefore, the steps to seek the appropriate transformation are the following.
First, we analyze the two-body correlation of the system.
Next, we invent a transformation which properly takes account of it.
When we use non-unitary transformations in this step, the non-Hermitian term
is not relevant as far as we concern low-energy excitations, 
and we can neglect it.
Finally, we obtain an effective Hamiltonian of the system.
Of course, the irrelevance of the non-Hermitian term must be shown
in a self-consistent way.

Now we remark on the relation to the usual Chern-Simons composite fermion
theory.
In the Chern-Simons formalism, the term which 
corresponds to $V^{NH}$ is a three-body interaction term.
When we perform the usual Chern-Simons singular gauge transformation,
the Hamiltonian is given by
\begin{equation}
H = H^0_{CS} + V^H_{CS} + V^3,
\label{cs_h}
\end{equation}
where $H^0_{CS}$ and $V^H_{CS}$ is given by replacing the 
$\pi_{\bf k}$ in Eq. (\ref{rs_h})
with $\phi^{\dagger}_{\bf k}$, which is the creation operator of 
composite particle with momentum ${\bf k}$ and
$V^3$ is the three-body interaction term for composite particles.
In the Chern-Simons gauge theory of the fractional quantum Hall effect
at $\nu=1/m$\cite{zhang}, where $m$ is an odd integer,
$V^3$ is irrelevant when the condensate of 
composite bosons occurs\cite{cb_three}.
We can apply this argument to the composite fermion pairing theory.
The pairings of composite fermions have the same role of composite bosons.
The power of the canonical dimension of the coupling constant for $V^3$ 
is half of that for composite bosons but the sign of it is unchanged.
Therefore, when the condensate of pairings of composite fermions occurs,
$V^3$ is no more relevant.

The effect of $V^3$ was discussed by Bonesteel \cite{bonesteel}
as the current-current correlation effect.
He concluded that it causes a pair-breaking effect.
The conclusion is similar to ours,
however, his argument that for the short range interaction 
this pair-breaking effect is strong enough to unstabilize the pairing state 
is not correct.
The pair-breaking effect caused by $V^3$ seems to exist but it is irrelevant 
for the pairing state.
If the pairing state is unstabilized, the pair-breaking effect is 
caused not from $V^3$ but
from the direct repulsive interaction between composite fermions 
or impurity potentials.

Here we remark on the Hamiltonian for the composite fermion theory.
When we deal with the pairing of composite fermions, 
we can neglect $V^{NH}$ as mentioned above. Therefore, it is enough to 
consider the Hamiltonian 
\begin{equation}
H=H^0 + V^H + V^C,
\end{equation}
for the pairing state of composite fermions.
However, when we deal with no-pairing states of composite fermions,
we cannot expect the irrelevance of $V^{NH}$ any more.
We must properly take into account the effect of $V^{NH}$, or $V^3$ for 
a compressible state of composite fermions.

This work is supported by Research Fellowships of the Japan Society for the 
Promotion of Science for Young Scientists.

\end{multicols}

\begin{references}
\bibitem{pairing} For parity and time reversal symmetry broken pairing 
states, it is possible to formulate it using the Chern-Simons gauge field.
The attractive interaction is caused from the minimal coupling between
the particles and the bare Chern-Simons gauge field 
(not the fluctuation of it).
\end{references}
\end{document}